\documentstyle[preprint,aps]{revtex}

\newcommand{ \BE }{ \begin{equation}}
\newcommand{ \EE }{ \end{equation} }

\newcommand{ \arch }{ \mbox{arch} }
\newcommand{\p }{ \varphi }
\newcommand{ \Rm }{ {\mathbf R} }

\newcommand{ \half }{ \frac{1}{2} }

\newcommand{ \kanskestorre}{ \stackrel{>}{_\sim} }
\newcommand{\std}{\langle\mbox{std}\ \p\rangle}

\title{Random background charges and Coulomb blockade in one-dimensional tunnel junction arrays}
\author{J. Johansson and D. B. Haviland}
\address{Royal Institute of Thecnology, Nanostructure Physics, 100 44 Stockholm, Sweden}
\begin{document}
\draft
\maketitle

\begin{abstract}
We have numerically studied the behavior of one dimensional tunnel junction arrays when random background charges are included using the ``orthodox'' theory of single electron tunneling. Random background charge distributions are verified in both amplitude and density. The use of a uniform array as a transistor is discussed both with and without random background charges. An analytic expression for the gain near zero gate voltage in a uniform array with no background charges is derived. The gate modulation with background charges present is simulated. 
\end{abstract}
\pacs{73.23.Hk,73.40.Gk}

\section{Introduction}

Since the discovery of correlated single electron tunneling many electronic devices exploiting this effect have been proposed~\cite{SCT,likharev_intro}. A gate is commonly used to control the device: In single electron transistors to modulate the current, in traps to pull electrons on to an island or in pumps and turn-styles to move electrons in a specific manner. The response to the gate is periodic in the induced charge on the gate capacitance, with a period of one electron. These devices are therefore very sensitive to charges trapped in the vicinity of the device, where there is a significant capacitive coupling to the islands. Randomly distributed ``background charge'' can change the threshold voltage for injecting charge in the array and the gate voltage where the coulomb blockade is lifted. The background charges may be static, causing a shift of the device operation point, or dynamic, causing noise in the device~\cite{zorin,starmark}.

The effect of background charges has been observed in experiments with small arrays. The influence of the background charge shows up as an offset in the current versus gate voltage curves~\cite{Q_coherence}. It has also been shown that the threshold voltage could be maximized by tuning gates to each island and thereby compensating for the background charge~\cite{Oudenaarden}. In single electron traps it was shown that the discrepancy between experiments and theory was likely to be due to random background charges~\cite{likharev_trapp}. Theoretical studies of the effect of random background charges in one dimensional arrays have examined how the threshold voltage is affected~\cite{Middleton,Melsen} and noise properties when background charges are included~\cite{likharev_noise}.

Here we discuss simulations of the transport properties of one dimensional arrays of series coupled tunnel junctions. The arrays are characterized by nearest neighbor capacitance $C$ and gate capacitance $C_g$. The simulations include random background charges which are characterized by their density and amplitude. We examine the effect of static background charges on the transport properties, including the gate voltage dependence, for various ratios $C/C_g$.

\section{Basic Equations}
\label{sec.BE}
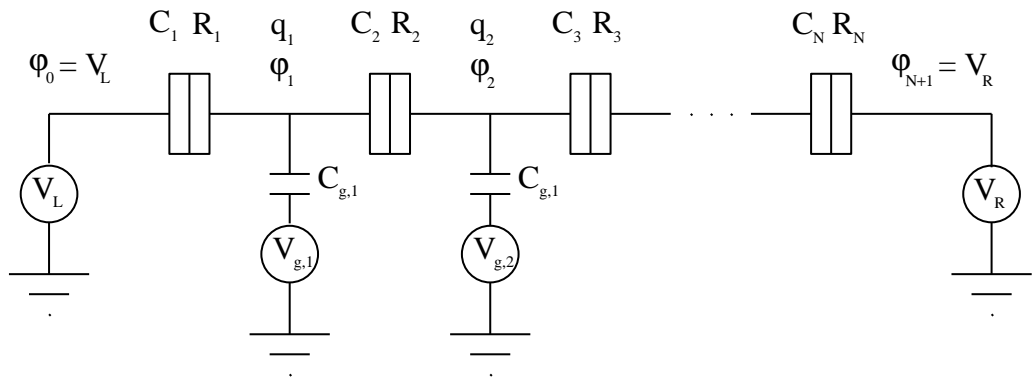
\begin{figure}
\caption{The schematic of a 1D array of small capacitance tunnel junctions.} 
\label{fig:array}
\end{figure}

The one dimensional array consists of $N-1$ small islands connected in series by small capacitance tunnel junctions. An equivalent circuit is shown in fig.~\ref{fig:array}. A gate voltage, $V_{g,i}$, is coupled through the gate capacitance $C_{g,i}$ to each island. A transport voltage is applied across the array by the external voltages $V_L$ and $V_R$. The tunnel junctions are characterized by the capacitance $C_i$ and the resistance $R_i$. If the tunnel resistance satisfies $R_i \gg  h/e^2 = 25.81281...\  \mbox{k}\Omega$, the number of charges on each island becomes sharply defined and the quasi-classical ``orthodox'' theory of single electron tunneling applies~\cite{SCT}. The theory predicts that electrons tunnel one at a time with a tunneling rate given by
\BE
\Gamma_i = \frac{1}{e^2R_i}\frac{\Delta F}{1-\exp(\Delta F/k_bT)}, \label{eq:rate}
\EE 
where $\Delta F$ is the difference in free energy of the initial and final state. We are interested in the change in free energy when one electron tunnels through one junction in the array, say from island $k$ to island $k\pm1$. In this case the free energy difference is given by the electrostatic potential difference of those islands before and after the tunneling event
\BE
\Delta F = F'-F = \frac{e}{2} [(\p_{k\pm 1}+\p'_{k\pm 1}) - (\p_k+\p'_k)]. \label{eq:dF}
\EE
In the case considered here, all junctions and gate capacitances are equal, $R_i = R$, $C_i = C$ and $C_{g,i} = C_g$. Charge conservation gives a relation between the potential on the islands and the charges on the islands:
\begin{eqnarray}
(2 + C_g/C)\p_1 - \p_2 &=& q_1/C + V_{L} \nonumber \\
-\p_{i-1} + (2 + C_g/C)\p_i - \p_{i+1} &=& q_i/C,\ \ \ \ 2 \leq i \leq N-2 \label{eq:rel} \\
-\p_{N-2} + (2 + C_g/C)\p_{N-1} &=& q_{N-1}/C + V_{R}. \nonumber
\end{eqnarray} 
where $q_i = en_i + q_{0,i} + en_g$ is the charge on island $i$, consisting of the induced background charge on that island $q_{0,i}$ and the induced gate charge $en_g$. The equations (\ref{eq:rel}) must be inverted to find the potential on the islands. This inversion can be done exactly~\cite{O'Connell}. The result of this inversion can be expressed in matrix form: $\overline{\p} = \Rm \overline{q}$, where the components of the symmetric matrix $\Rm$ are
\BE
\Rm_{ij} = \frac{\cosh(N-|i-j|)\lambda-\cosh(N-i-j)\lambda}{2\sinh\lambda \sinh N\lambda} \label{eq:matrix}
\EE
and $\lambda = \arch(1 + C_g/2C)$ is known as the inverse of the charge soliton length introduced in ref.~\cite{likharev_mc}. The charge soliton length, $\lambda^{-1}$, is the screening length giving the distance over which the potential from one excess electron is screened by the free charge in the ground plane. With equations~(\ref{eq:rate})-(\ref{eq:matrix}), all tunneling rates can be calculated. A Monte Carlo scheme~\cite{likharev_mc} is used for the simulation of the flow of charges. The random numbers used to generate the background charge configuration and for the monte Carlo calculation are chosen to make a good quasi random series~\cite{C}. All results are calculated for $T=0$. 

\section{Effect of random background charges}

Charges trapped in the substrate or in the tunnel barriers induce an excess charge on the islands, $q_i \rightarrow q_i + q_{0i}$.  The background charge $q_{0i}$, is between -e/2 and e/2 because any value outside this interval would be compensated for by one electron or hole tunneling on to the island. 

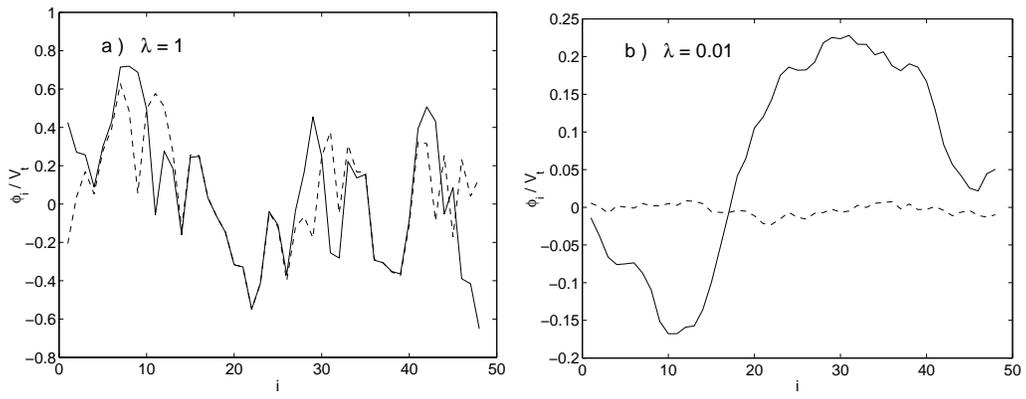
\begin{figure}
\caption{The potential in a 50 junction array as a function of the island number for $\lambda = 1$ and $\lambda = 0.01$. The solid line is before the free energy has been minimized by single charge tunneling, and the dashed line is after minimization.} 
\label{fig:relaxed}
\end{figure}

Starting from a random background charge configuration, under certain conditions tunneling events will lower the free energy by adding or rearranging electrons or holes. Since the electrostatic energy is proportional to the gradient of the potential squared, lowering the energy means smoothing the potential.  This possibility to relax the random potential variations depends on the length of the array, the soliton length, and on the actual configuration of the background charges. In fig.~\ref{fig:relaxed} the potential before and after relaxation is shown for two different values of $\lambda$. One can see by looking at fig.~\ref{fig:relaxed}a that for short soliton length the potential is nearly unchanged by the relaxation process. However for long soliton length (fig.~\ref{fig:relaxed}b) we can see how the relaxation reduces variation of the potential. The potentials plotted in figure 2 for different $C/C_g$ are normalized by the threshold voltage. The threshold voltage (derived below) is the proper measure of the Coulomb blockade for transport in an array. 

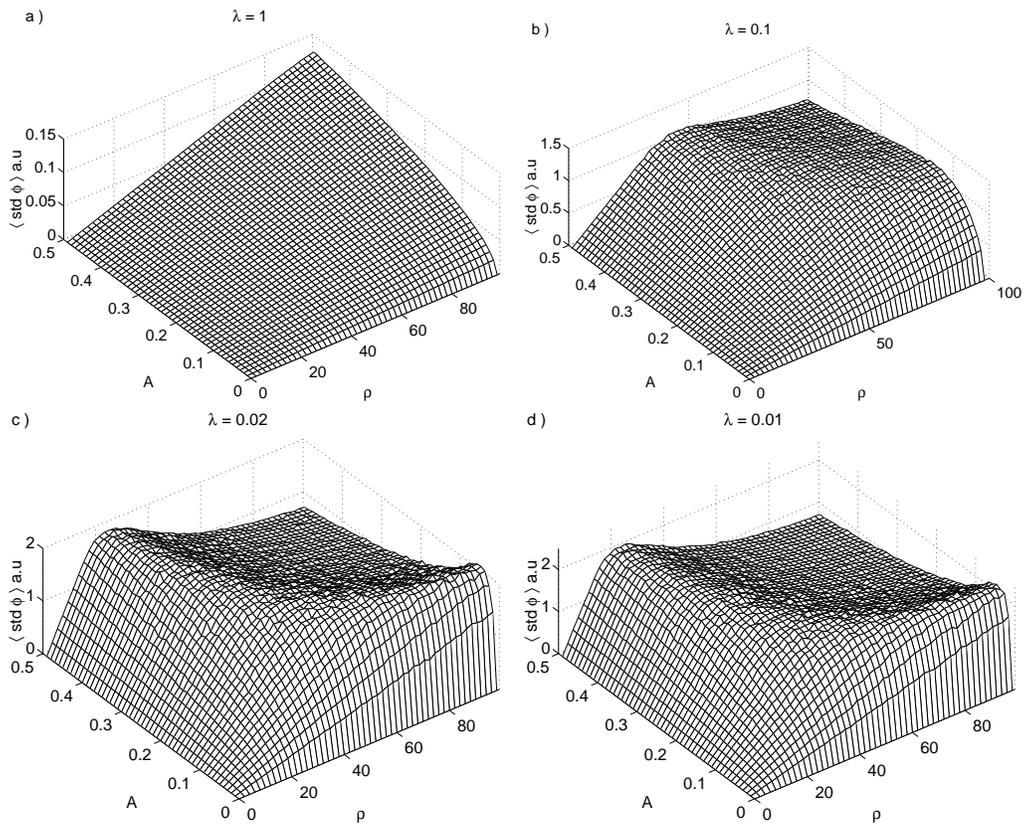
\begin{figure}
\caption{The mean value of the standard deviation of the potential in the array of 100 junctions for different values of $\lambda$.} 
\label{fig:3D}
\end{figure}

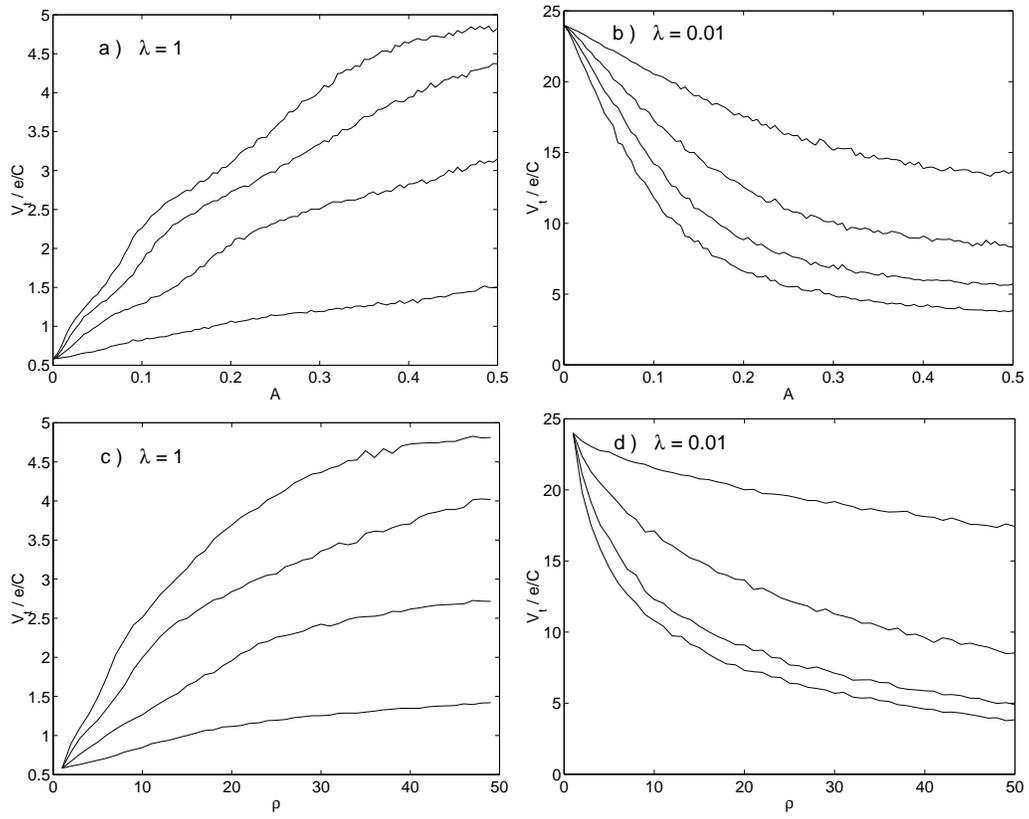
\begin{figure}
\caption{The threshold voltage in a 50 junction array as a function of the amplitude, $A$ (left panels) and the density,$\rho$ (right panels) of the background charges. The different curves on the left correspond to $\rho =$ 49, 30, 15 and 5.  The different curves on the right correspond to $A =$ 1, 0.6, 0.3 and 0.1.}
\label{fig:Vt_A_r}
\end{figure}

The random background charge configuration can be described by two parameters: the number of islands which have an induced charge $q_{0,i} \not = 0$, or the density of induced charges $\rho \in \{0, N-1\}$, and the amplitude $A$ of the induced charges $q_{0,i} \in \{-Ae, +Ae\}$ with $0 \leq A \leq 0.5$. In fig.~\ref{fig:3D} the mean value of the ``relaxed'' potential variation along the array $\std$, averaged over 1000 configurations of random background charges, is plotted versus $\rho$ and $A$. The four plots in fig.~\ref{fig:3D} correspond to four different values of the soliton length $\lambda^{-1}$. Here we can see that the smoothing of the potential becomes more effective when the soliton length increases. Comparison of figures~\ref{fig:3D}c and ~\ref{fig:3D}d show that when $\lambda^{-1} \kanskestorre N$ the value of $\std$ is weakly dependent on $\lambda$. Further we can see that when $\lambda^{-1} \kanskestorre N$ and $\rho \kanskestorre N/2$, $A \kanskestorre 0.25$, the potential variations of the array are weakly dependent on $A$ and $\rho$. This weak dependence can also be seen in fig.~\ref{fig:Vt_A_r} where the threshold voltage is plotted as a function of the amplitude (right panels) and density (left panels) for various $\lambda$. The threshold voltage changes almost linearly with the amplitude and density of the background charges when $\lambda = 1$ (figs. \ref{fig:Vt_A_r}a and \ref{fig:Vt_A_r}c), but when $\lambda = 0.01$ the dependence is very weak for $A \kanskestorre 0.25$ and/or $\rho \kanskestorre N/2$ (figs. \ref{fig:Vt_A_r}b and \ref{fig:Vt_A_r}d). Thus we see that as far as the electrostatics is concerned, disorder due to random background charges may be effectively reduced by single charge tunneling if the soliton length is larger than the array length and the density and amplitude of random background charges is large enough.

\section{Gate effects in 1D arrays}

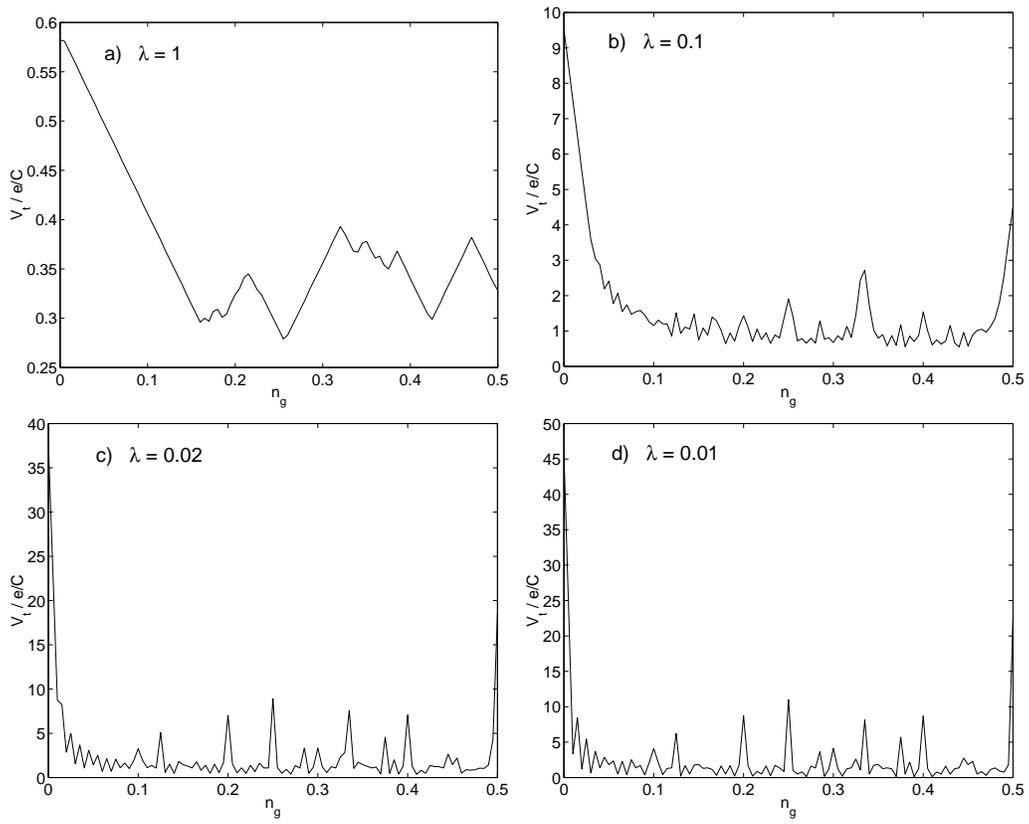
\begin{figure}
\caption{The effect of the gate on the threshold voltage for an array with no background charges present, N = 100.} 
\label{fig:gate_mod}
\end{figure}

When a gate voltage is capacitively coupled to the islands of the array the current voltage characteristics are strongly affected around the threshold. In the case when the gate is uniformly applied ($V_{g,i} = V_g$ and $C_{g,i} = C_g$) and there are no background charges, the threshold voltage changes with induced charge as shown in fig.~(\ref{fig:gate_mod}). For appropriate values of $N$ and $\lambda$ an array with uniform coupled gate capacitance has ideal transistor characteristics. Such an array would make a good switch because of the initial steep dependence of $V_t$ on $n_g$, and the wide region of $n_g$ with a weaker dependence. 

The sensitivity to gate voltage is greatest when $N$ is large but $N < \lambda^{-1}$. In this case the threshold voltage is large at zero gate voltage~\cite{O'Connell}, and the gate drives the threshold almost to zero. The initial linear decrease of $V_t$ with $n_g$ can be examined analytically because during this decrease the gate voltage is not large enough to bind any charges inside the array. When no charges are localized in the array (electrons, holes or background charges) it was shown in ref.~\cite{O'Connell} that the threshold voltage is determined by the tunneling rates through the two end junctions. With eq.~(\ref{eq:matrix}) we find the relevant potentials to calculate the tunneling rate through the end junction
\BE
\p_1 = \frac{e}{C}\left[\Rm_{1,1}(1 + CV_L/e) + n_g\sum_{j=1}^{N-1}\Rm_{1,j}\right],
\EE
\BE
\p'_1 =\frac{e}{C}\left[\Rm_{1,1}V_LC/e + n_g\sum_{j=1}^{N-1}\Rm_{1,j}\right],
\EE
\BE
\p_0 = \p'_0 = V_L.
\EE
Here we have taken $V_L > 0$, $V_R = 0$ and $V_g < 0$ so in this case a hole will tunnel in from the left. Making use of the relations 
\BE
\Rm_{1,j} = \frac{\sinh(N-j)\lambda}{\sinh(N\lambda)},
\EE
\BE
\sum_{j=1}^{N-1}\Rm_{1,j} = \frac{\sinh[\half(N-1)\lambda]}{2\cosh(N\lambda/2)\sinh(\lambda/2)}.
\EE
We obtain the tunneling rate from eq.~(\ref{eq:dF}). Setting the tunneling rate to zero and solving for $V_L$ we get the threshold voltage:
\BE
V_t = \frac{e}{C}\left[\frac{\sinh(N-1)\lambda}{\sinh N\lambda - \sinh(N-1)\lambda}- \frac{\sinh[\half(N-1)\lambda] \sinh N\lambda/2}{\sinh^2\lambda/2 \cosh(N-\half)\lambda} n_g \right]. \label{eq:V_t}
\EE
If $n_g$ is set to zero in this expression we recover the expression for the threshold voltage found in~\cite{O'Connell}. This equation explains the very steep slope of the decrease in threshold voltage, for large $\lambda^{-1}$ and $N$. 

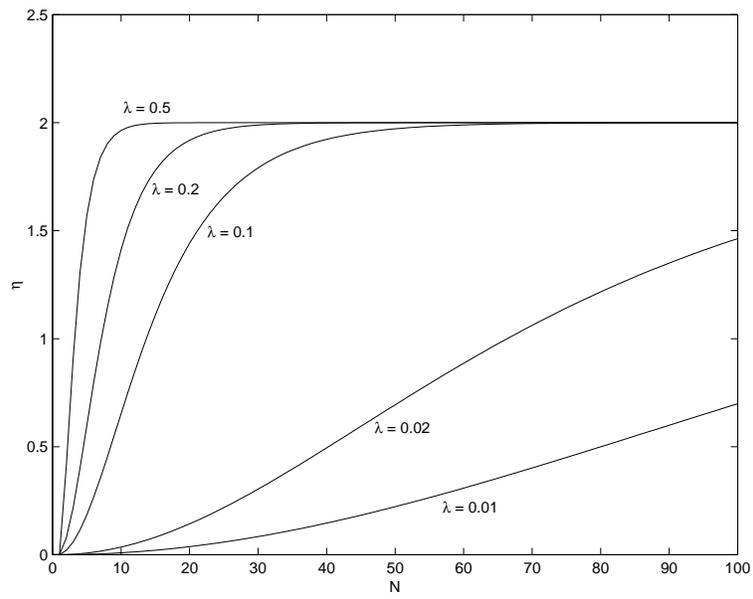
\begin{figure}
\caption{The voltage gain near zero gate voltage of an array plotted veurses $N$ for different values of $\lambda$. The analytic expresion is given in equ. \ref{eq:gain} valid for arbitrary $N$ and $\lambda$.} 
\label{fig:gain}
\end{figure}

Eq.~(\ref{eq:V_t}) can be used to calculate the voltage gain of the uniform array. If the array is current biased very close to zero and it is assumed that the gate voltage drives the threshold voltage to zero, the voltage gain is given by
\BE
\eta = \frac{\mbox{d}V_{out}}{\mbox{d}V_g} \approx \frac{C_g}{e}\frac{\mbox{d}V_t}{\mbox{d}n_g} = 4 \frac{\sinh[\half(N-1)\lambda] \sinh N\lambda/2}{ \cosh(N-\half)\lambda}. \label{eq:gain}
\EE
The gain is plotted in fig.~\ref{fig:gain} for different values of $\lambda$. The gain approaches 2 when $\lambda N \kanskestorre 3$, which resembles the voltage gain in the uniform single electron transistor where one needs $C_g \gg C$ to get a maximum gain of 2.

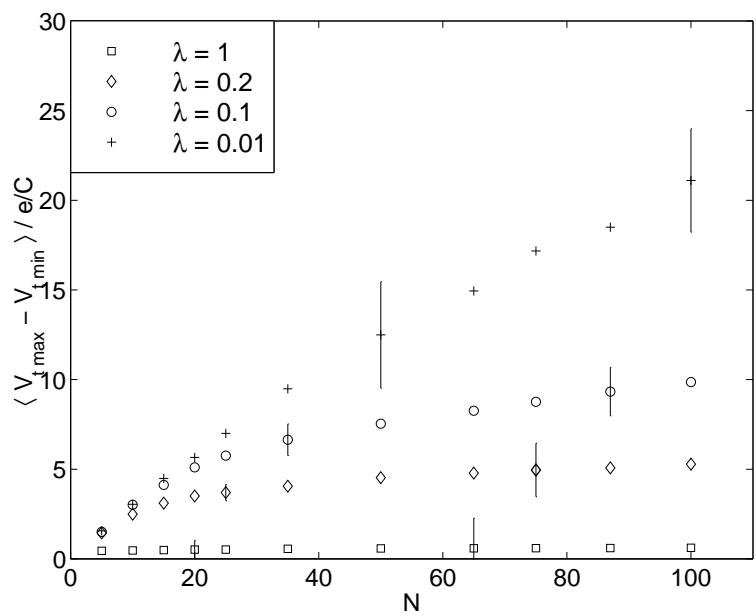
\begin{figure}
\caption{The averagy modulation of the threshold voltage plotted versus $N$ for various $\lambda$. The errorbars give the standard deviation calculated from 1000 random background charge configurations.} 
\label{fig:modulation}
\end{figure}

Unfortunately this gain behavior near zero gate voltage is destroyed by random background charges. Melsen~{\it et.~al.}~\cite{Melsen} have shown that the mean value (averaged over background charge configurations) of the threshold voltage was independent of the gate voltage. Our simulations show that it is still possible to modulate the threshold voltage with gate voltage when background charges are present, however the modulation changes. The large peak and high sensitivity around zero gate voltage disappears, and the largest modulation can occur near any value of $n_g$. The maximum modulation ($V_t|_{max} - V_t|_{min}$) also becomes smaller than in the no-background-charge case.  In fig.~\ref{fig:modulation} the maximum modulation (averaged over 1000 background charge configurations) is plotted as a function of number of junctions for different values of $\lambda$. As in the no-background-charge case, the modulation becomes larger with increasing $N$ and $\lambda^{-1}$. Here again, we see weak dependence of the modulation on $N$ when $N\lambda^{-1} > 1$. The error bars plotted for several points give the standard deviation of the modulation, or the spread in modulation depending on the particular configuration of background charges. Thus, static random background charges cause random shift in the operation point in gate voltage, reduce the maximum modulation with gate voltage, and cause significant variation in the gate modulation of the array.

We have simulated the effects of random background charges on the Coulomb blockade in one dimensional tunnel junction arrays. We find that when $\lambda^{-1} \kanskestorre N$ the static potential along the array is smooth and does not depend very strongly on the density or amplitude of background charges. We derive an analytical expression for the voltage gain of an array transistor, valid when $n_g$ is small and no background charges present. However, by simulating the behavior of the threshold voltage as a function of a symmetrically coupled gate voltage with background charges in the array, we conclude that random background charges render arrays useless in transistor applications.

We gratefully acknowledge financial support from the Swedish TFR/NFR and the European Union through MEL ARI project 22953 CHARGE.

\end{document}